\documentclass{aastex}          
\usepackage{spr-astr-addons}    
\usepackage{natbib}             
\usepackage{multirow}
\usepackage[hyperindex,breaklinks]{hyperref}

\newcommand{\tens}{\mathbf}
\begin{document}

\title{Yet another sample of RFGC galaxies}

\shorttitle{Yet another sample of RFGC galaxies}
\shortauthors{S.~L.~Parnovsky \and A.~S.~Parnowski}

\author{S.~L.~Parnovsky} 
\affil{Astronomical Observatory, Taras Shevchenko National University of Kyiv\\
Observatorna str. 3, 04058 Kyiv, Ukraine\\
tel: +380444860021, fax: +380444862191\\ e-mail:par@observ.univ.kiev.ua}
\email{par@observ.univ.kiev.ua}

\author{A.~S.~Parnowski} 
\affil{Space Research Institute\\
prosp. Akad. Glushkova 40 build. 4/1, 03680 MSP Kyiv-187, Ukraine\\
tel: +380933264229, fax: +380445264124\\ e-mail:parnowski@ikd.kiev.ua}
\email{parnowski@ikd.kiev.ua}

\begin{abstract}
We present a new version of a sample of galaxies from the Revised Flat Galaxy 
Catalogue (RFGC), which have redshift and \mbox{H\,{\sc i}} line width data. We 
also give the parameters of the collective motion model determined upon this 
sample. The considered models of motion include the dipole (bulk flow), the 
quadrupole (cosmic shear) and the octupole components. We also considered 
higher-order multipoles. In all cases the obtained parameters matched the 
$\Lambda$CDM cosmology.
\end{abstract}


\section{Introduction}\label{s:Introduction}
The distribution of matter density is inhomogeneous in the region of the 
Universe, limited to about $100h^{-1}\,\textrm{Mpc}$, which contains several 
superclusters and voids. A galaxy, besides the cosmological expansion, is also 
attracted to the regions with greater density. Due to this fact, the galaxies 
are involved in a large-scale collective motion on the background of Hubble 
expansion. Investigation of such a motion is important since it allows plotting 
the distribution of matter in the surrounding region of the Universe and 
comparing this distribution with the distribution of luminous matter. 

This is especially important due to the fact that the most serious challenges 
to the standard $\Lambda$CDM cosmology are posed by the inconsistencies in the 
estimations of the velocity of the bulk motion. The $\Lambda$CDM model 
estimates it at the level $\sim 250\,\textrm{km\,s}^{-1}$ at the scale 
$100h^{-1}\,\textrm{Mpc}$. However, in some studies, for example 
\citet{ref:LP,ref:WFH09,ref:FWH10}, the obtained values were larger by a factor 
of 2. The largest value of $416\pm 79\,\textrm{km\,s}^{-1}$ was given by 
\citet{ref:FWH10}. Other results, including our own, give smaller values, which 
are consistent with the predictions of the $\Lambda$CDM cosmology.

In addition to redshifts we require independent estimations of distances to the 
galaxies. Since we deal only with spiral galaxies, we use the Tully-Fisher 
relation to determine these distances. In its common version it relates the 
intrinsic luminosity of a galaxy with its velocity width (the amplitude of its 
rotation curve). However, we use the `\mbox{H\,{\sc i}} line width -- linear 
diameter' variant of the Tully-Fisher relation, which does not require 
photometric data. It is different from the common version and the data are also 
processed in a different fashion. If these two variants give similar results, 
this is a strong evidence of their correctness. For this reason, it is 
important to continue the research of the collective motions of galaxies, even 
if it can not boast the best accuracy or exceptionally large depth. Of course, 
different versions of the Tully-Fisher relation can not be regarded as 
independent methods, but they can potentially give very different results.

Due to large errors in determination of distances to galaxies, which are caused 
both by measurement errors and by the intrinsic uncertainty of the Tully-Fisher 
relation, it is necessary to compile large samples, and to pay special 
attention to the adequate processing of the data. Also, the choice of the model 
of the collective motion strongly affects the results. For low-depth samples it 
was possible to use the simple bulk motion model. However, this model is 
inadequate for deeper samples and more complex models should be used. These 
models contain higher-order multipoles and take into account additional 
effects. 

We study these motions using the Revised Flat Galaxy Catalogue 
\citep[RFGC,][]{ref:RFGC}. It contains data about $N=4236$ galaxies including 
the information on the following parameters: Right Ascension and Declination 
for the epochs J2000.0 and B1950.0, galactic longitude and latitude, major and 
minor blue and red diameters in arcminutes in the POSS-I diameter system, 
morphological type of the spiral galaxies according to the Hubble 
classification, index of the mean surface brightness (I -- high, IV -- very 
low) and some other parameters, which are not used in this article. More 
detailed description of the catalogue can be found in the paper 
\citep{ref:RFGC}.

\section{Data used}\label{s:data}
We used the sample containing 1720 RFGC galaxies. The data for 59 galaxies were 
rejected due to large deviations from the Tully-Fisher relation. Thus, the 
resulting sample contained 1661 galaxies. In addition to RFGC data we used the 
data on the radial velocities of the galaxies and \mbox{H\,{\sc i}} line width 
at the $50\%$ level. As usual, all velocities were converted to the CMB frame 
and \mbox{H\,{\sc i}} line widths were corrected for intrinsic absorption and 
turbulence. The new data were taken from the SFI++ II survey 
\citep{ref:SFI++2}, 3 releases of the ALFALFA survey 
\citep{ref:A3,ref:A6,ref:A8} and the 40ALFA survey \citep{ref:A40}, which 
contains about $40\%$ of data to enter the final version of the ALFALFA survey. 
Some results of RFGC galaxies' observation at the Effelsberg radio telescope 
were taken from the papers \citep{ref:Mitr05,ref:Kudrya09}. From all other 
sources only one measurement was included in this sample from the paper 
\citep{ref:Kovac09}, which contains the results of a blind survey in the Canes 
Venatici region. Thus, the new sample is based on much more homogeneous data 
than the previous one, which is an additional benefit.

As a result, we added the data about 42 new galaxies, 5 of which were later 
rejected. In addition, the data on a large number of galaxies, present in the 
previous version of the sample, were re-measured. The majority of such data 
featured only slight modifications with respect to the previous version. This 
suggests that these data are of good quality. Nevertheless, we changed the data 
for 190 out of 1623 galaxies from the previous sample. For 88 galaxies the 
changes were minor and did not drastically affect the results. However, the 
data for 47 galaxies were changed significantly. This means that for these 
galaxies either the \mbox{H\,{\sc i}} line width changed by more than 
$20\,\textrm{km\,s}^{-1}$ or the radial velocity changed by more than 
$40\,\textrm{km\,s}^{-1}$, the prior being more common. The maximal change of 
velocity exceeded $2000\,\textrm{km\,s}^{-1}$ and the maximal change of 
\mbox{H\,{\sc i}} line width was $126\,\textrm{km\,s}^{-1}$. We will give the 
details of the sample construction in Section \ref{s:results}.

Thus, we can say that our sample has 37 added + 47 replaced = 84 essentially 
different data, which equals to $5\%$ of the sample volume. This increase of 
the volume is smaller than in the previous versions of the sample. This is 
likely due to the fact that most galaxies in the field of regard of the Arecibo 
radio telescope were already measured and the observations in the southern 
hemisphere are lagging behind. The progress in the Arecibo zone can be provided 
by the remaining $60\%$ of the ALFALFA survey and the improved measurements of 
the rejected galaxies. The greatest impact would be provided by the southern sky 
observations, since these are quite scarce and the addition of new data would 
not only increase the sample volume but also make it more uniform and symmetric. 

\section{Models of the collective motion}\label{s:models}
In our previous articles \citep{ref:Par01,ref:ParTug04,ref:APSS10,ref:APSS11} 
we described the models of the collective motion we used. Here we will only 
give two models given there. We will start from the most complex of the 
previously considered models, namely the DQO-model. 
\begin{equation}\label{eqn:DQO}
V=R+V^{dip}+V^{qua}+V^{oct}+\Delta V.
\end{equation}
Here $V$ is a radial
velocity of the galaxy in the CMB isotropy frame, $R=Hr$ is the Hubble
velocity, $r$ is the distance to the galaxy, $\Delta V$ is a random error,
$V^{dip}$, $V^{qua}$ and $V^{oct}$ are the dipole (D), quadrupole (Q) and
octupole (O) components of the non-Hubble cosmic flow. They are given by the
following expressions:
\begin{equation}\label{eqn:VDQO}
\begin{array}{l}
V^{dip}=D_{i}n_{i},\\
V^{qua}=RQ_{ik}n_{i}n_{k}\\
\phantom{V^{qua}}{}=R\left(q_1(n_1^2-n_3^2)+q_2(n_2^2-n_3^2)\right.\\
\phantom{V^{qua}}\left.{}+q_3n_1n_2+q_4 n_1 n_3+q_5 n_2 n_3\right),\\
V^{oct}=R^{2}(O_{ikl}n_{i}n_{k}n_{l}+P_{i}n_{i})=R^2\left(P_{i}n_{i}\right.\\
\phantom{V^{oct}}\left.{}+o_1(3n_1n_2^2-n_1^3)+o_2(3n_1n_3^2-n_1^3)\right.\\
\phantom{V^{oct}}\left.{}+o_3(3n_2n_1^2-n_2^3)+o_4(3n_2n_3^2-n_2^3)\right.\\
\phantom{V^{oct}}\left.{}+o_5(3n_3n_1^2-n_3^3)+o_6(3n_3n_2^2-n_3^3)\right.\\
\phantom{V^{oct}}\left.{}+o_7n_1n_2n_3\right).
\end{array}
\end{equation}
Here we use Einstein's convention -- summation by repeated indices. $n_i$ are 
the Cartesian components of the unit vector $\vec{n}$ towards the galaxy, 
connected with the galactic coordinates $l$ and $b$ in the following way:
\begin{equation}\label{eqn:lb}
\begin{array}{l}
n_1=n_z=\sin b,\\
n_2=n_x=\cos l\cos b,\\
n_3=n_y=\sin l\cos b.
\end{array}
\end{equation}
The dipole component (bulk motion) is described by the vector $\vec{D}$. The 
quadrupole component (cosmic shear) is described by the symmetrical traceless 
tensor $\tens{Q}$. It has 5 independent parameters $q_i$. The octupole 
component can be described by one rank 3 tensor, but we divide it into a trace 
characterised by vector $\vec{P}$ and a tensor $\tens{O}$, which is 
antisymmetrical with respect to each pair of indices. The latter has 7 
independent parameters $o_i$.

Hubble velocity is determined from the generalised Tully-Fisher relation in the
``angular diameter -- \mbox{H\,{\sc i}} line width'' version by the following 
formula 
\begin{equation}\label{eqn:TF}
\begin{array}{l}
R=(C_1+C_2B+C_3BT+C_4U)\frac{W}{a}\\
\phantom{R=}{}+C_5\left(\frac{W}{a}\right)^2+C_6\frac{1}{a},
\end{array}
\end{equation}
where $W$ is the corrected \mbox{H\,{\sc i}} line width in 
$\textrm{km\,s}^{-1}$ measured at $50 \%$ of the maximum, $a$ is the corrected 
major galaxies' angular diameter in arcminutes on red POSS and ESO/SERC 
reproductions, $U$ is the ratio of major galaxies' angular diameters on red and 
blue reproductions, $T$ is the morphological type indicator ($T=I_{t}-5.35$, 
where $I_{t}$ is Hubble type; $I_{t}=5$ corresponds to type Sc), and $B$ is the 
surface brightness indicator ($B=I_{SB}-2$, where $I_{SB}$ is the surface 
brightness index from RFGC; brightness decreases from I to IV). Note that the 
statistical significance of each term in eq. (\ref{eqn:TF}) is greater than 
$99\%$ according to the F-test \citep{ref:F,ref:H}.

Thus, the DQO-model contains 24 parameters, namely 3 components of the vector
$\vec{D}$, 6 coefficients $C_i$, 5 parameters $q_i$ of the tensor $\tens{Q}$, 3
components $p_i$ of the vector $\vec{P}$ and 7 parameters $o_i$ of the tensor
$\tens{O}$. By rejecting $V^{oct}$ we get a simpler DQ-model with 14 
components. Further rejecting $V^{qua}$ leads to the simplest D-model with 9 
components.

In the article \citep{ref:APSS11} on the base of the results of 
\citet{ref:KudAlex02,ref:KudAlex04} we also introduced relativistic models of 
motion based on the idea that for the homogeneous isotropic cosmological models 
the dependence of the velocity $V=cz$ on the angular diameter distance $R=Hr$ 
in the next order in $R$ has the form
\begin{equation}\label{eqn:R}
V=R+\gamma R^2.
\end{equation}
The coefficient $\gamma$ is expressed through the deceleration parameter $q$ by
\begin{equation}\label{eqn:g}
\gamma=\frac{3+q}{2c}.
\end{equation}
For the standard $\Lambda$CDM cosmology we have
\begin{equation}\label{eqn:q}
q=\frac{\Omega_m}{2}-\Omega_{\Lambda}=-0.61,
\end{equation}
where $\Omega_m$ and $\Omega_{\Lambda}$ are the relative densities of matter,
including dark matter, and dark energy respectively. Numerical estimations are
based on the results of 7-year WMAP observations \citep{ref:WMAP7}. Introducing
(\ref{eqn:q}) into (\ref{eqn:g}) we obtain
\begin{equation}\label{eqn:g0}
\gamma_0=3.98\cdot 10^{-6}\,\textrm{s\,km}^{-1}.
\end{equation}

In this article we use this so-called semi-relativistic model of galaxy motion 
with fixed $\gamma$ in the inhomogeneous space-time, which has the form 
\begin{equation}\label{eqn:DQOR}
V_{rel}=R+V^{dip}+V^{qua}+V^{oct}+\gamma_0 R^2+\Delta V.
\end{equation}
The use of $\gamma_0$ mitigates the negative impact of measurement errors in 
the \mbox{H\,{\sc i}} line widths and the angular diameters. Please refer to 
the articles \citep{ref:JPS10,ref:APSS11} for details. 
Since we consider the terms proportional to
$R^2$ separately, we should remove the terms quadratic in distance from the
generalised Tully-Fisher relation (\ref{eqn:TF}):
\begin{equation}\label{eqn:TFR}
R=(C_1+C_2B+C_3BT+C_4U)\frac{W}{a}+C_5\frac{1}{a}.
\end{equation}
Note that all the remaining terms in this equation are inverse proportional to
the angular diameter $a$.

It is possible to add the next two multipoles to the DQO-model by constructing 
the DQOX-model 
\begin{equation}\label{eqn:DQOX}
V=R+V^{dip}+V^{qua}+V^{oct}+V^{hex}+\Delta V
\end{equation}
and the DQOXT-model
\begin{equation}\label{eqn:DQOXT}
V=R+V^{dip}+V^{qua}+V^{oct}+V^{hex}+V^{(32)}+\Delta V.
\end{equation}
The velocity components corresponding to the 16-pole and the 32-pole have the 
form 
\begin{equation}\label{eqn:XT}
\begin{array}{l}
V^{hex}=R^{3}X_{ijkl}n_{i}n_{j}n_{k}n_{l}=R^3\left(x_1n_1^4 + x_2n_2^4 \right.\\
\phantom{V^{hex}}\left.{}+ x_3n_3^4 + x_4n_1^3n_2 + x_5n_1^3n_3 + x_6n_2^3n_1 \right.\\
\phantom{V^{hex}}\left.{}+ x_7n_2^3n_3 + x_8n_3^3n_1 + x_9n_3^3n_2 + x_{10}n_1^2n_2^2 \right.\\
\phantom{V^{hex}}\left.{}+ x_{11}n_1^2n_3^2 + x_{12}n_3^2n_1^2 + x_{13}n_1^2n_2n_3 \right.\\
\phantom{V^{hex}}\left.{}+ x_{14}n_2^2n_1n_3 + x_{15}n_3^2n_1n_2 \right),\\
V^{(32)}=R^{4}T_{ijklm}n_{i}n_{j}n_{k}n_{l}n_{m} = R^4\left(t_1n_1^5 \right.\\
\phantom{V^{(32)}}\left.{}+ t_2n_2^5 + t_3n_3^5 + t_4n_1^4n_2 + t_5n_1^4n_3 \right.\\
\phantom{V^{(32)}}\left.{}+ t_6n_2^4n_3 + t_7n_2^4n_1 + t_8n_3^4n_1 + t_9n_3^4n_2 \right.\\
\phantom{V^{(32)}}\left.{}+ t_{10}n_1^3n_2^2 + t_{11}n_1^3n_3^2 + t_{12}n_2^3n_3^2 \right.\\
\phantom{V^{(32)}}\left.{}+ t_{13}n_2^3n_1^2 + t_{14}n_3^3n_1^2 + t_{15}n_3^3n_2^2 \right.\\
\phantom{V^{(32)}}\left.{}+ t_{16}n_1^3n_2n_3 + t_{17}n_2^3n_3n_1 + t_{18}n_3^3n_1n_2 \right.\\
\phantom{V^{(32)}}\left.{}+ t_{19}n_1^2n_2^2n_3 + t_{20}n_2^2n_3^2n_1 + t_{21}n_3^2n_1^2n_2\right).
\end{array}
\end{equation}

However, it is more correct to introduce these multipoles to the 
semi-relativistic model of motion 
\begin{equation}\label{eqn:DQOXR}
V_{rel}=R+V^{dip}+V^{qua}+V^{oct}+V^{hex}+\gamma_0 R^2+\Delta V
\end{equation}
and
\begin{equation}\label{eqn:DQOXTR}
V_{rel}=R+V^{dip}+V^{qua}+V^{oct}+V^{hex}+V^{(32)}+\gamma_0 R^2+\Delta V,
\end{equation}
where $R$ is given by the relativistic Tully-Fisher relation (\ref{eqn:TFR}).
This is due to the $C_5\left(\frac{W}{a}\right)^2$ term in the non-relativistic 
Tully-Fisher relation (\ref{eqn:TF}), which causes different multipoles to be 
indistinguishable. 

\section{Results}
We process the new sample taking into account each model of motion described 
above. We created subsamples with a limitation on the distance in the 
non-relativistic D-model $R < R_{max}$. This is the most straightforward way 
for limiting the sample depth. The mean distances for the two most often used 
subsamples with $R_{max} = 80 h^{-1}\,\textrm{Mpc}$ and $R_{max} = 100 
h^{-1}\,\textrm{Mpc}$ are equal to $45.4 h^{-1}\,\textrm{Mpc}$ and $51.5 
h^{-1}\,\textrm{Mpc}$ respectively. For each subsample we determined the
regression coefficients using the least square method simultaneously in the 
Tully-Fisher relation and the model of motion. The results obtained allow 
estimating the quality of the sample.

\begin{figure}[tb]
\includegraphics[width=\columnwidth]{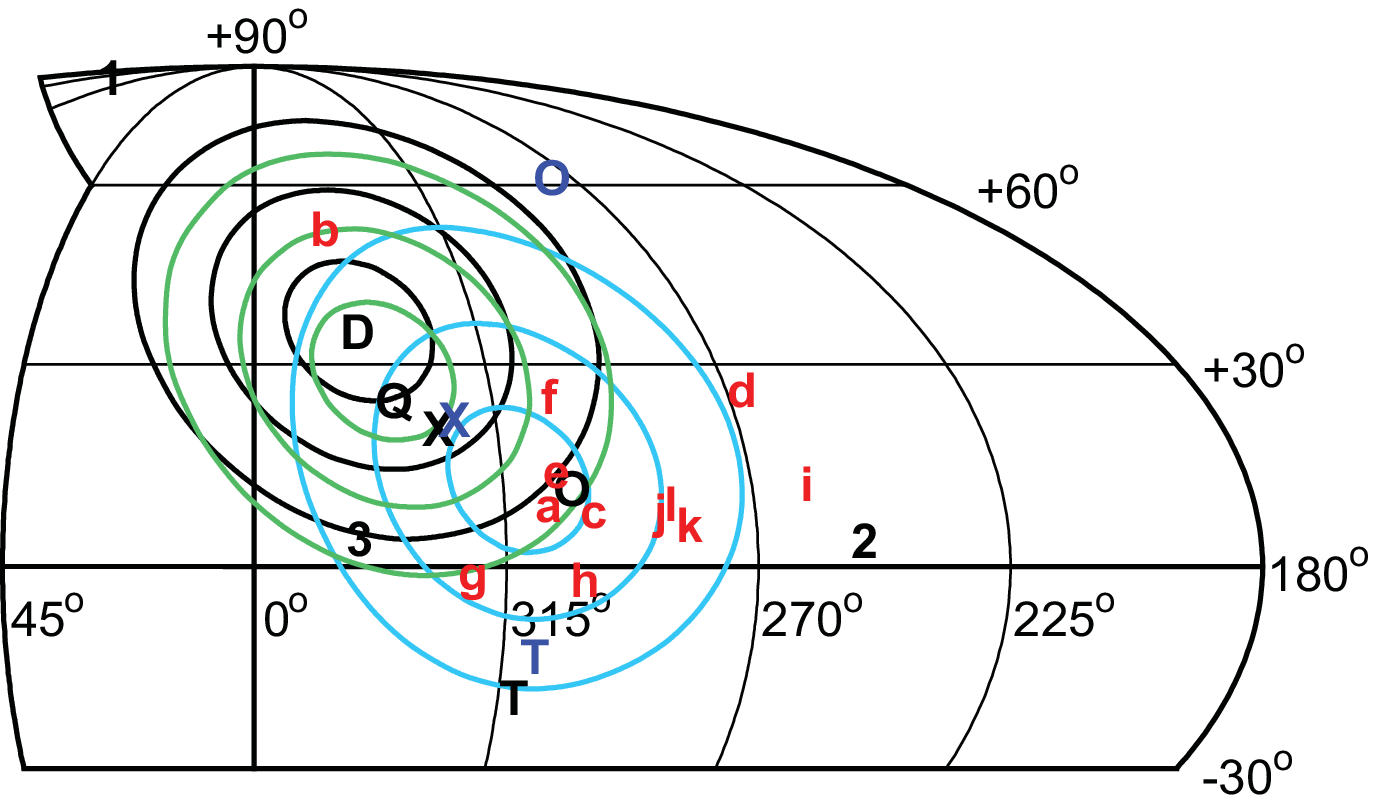}
\caption{A fragment of the Mollweide projection of the celestial sphere.
Solid lines show the $1\sigma$, $2\sigma$ and $3\sigma$ confidence boundaries 
of the bulk flow apices at $100 h^{-1}\,\textrm{Mpc}$. Black -- the new sample, 
green -- \citep{ref:APSS10}, sky blue -- \citep{ref:ParTug04}.
The results of other authors are marked with red lowercase letters.
a -- \citep{ref:LB}, b -- \citep{ref:LP}, c -- \citep{ref:H95}, d -- 
\citep{ref:Dale99}, e -- \citep{ref:Dekel99}, f -- \citep{ref:daCosta00}, g -- 
\citep{ref:Par01}, h -- \citep{ref:Kudrya03}, i -- \citep{ref:SMAC04}, j -- 
\citep{ref:WFH09}, k -- \citep{ref:FWH10}, l -- the composite sample from 
\citep{ref:Ma11}.
The apices in different models of motion are shown with uppercase letters.
D -- D-model, Q -- DQ-model, O -- DQO-model, X -- DQOX model, T -- DQOXT-model.
For DQO-, DQOX- and DQOXT-models the non-relativistic versions are shown with 
black colour and the semi-relativistic -- with blue colour.
The numbers show the directions of the quadrupole eigenvectors in the 
DQO-model. 1 -- maximum, 2 -- intermediate, 3 -- minimum.}
\label{fig:1}
\end{figure}

The main change since the previous sample 
is the slightly different apex of the dipolar component. This is demonstrated 
in Figure \ref{fig:1} where the boundaries of $1\sigma$, $2\sigma$ and 
$3\sigma$ confidence areas of the apices of the dipolar component are shown for 
3 different samples using the D-model with $R_{max} = 100 
h^{-1}\,\textrm{Mpc}$. The new sample is shown with black lines, the sample 
introduced in the articles \citep{ref:arxiv09,ref:APSS10} is shown with light 
green lines, and the sample introduced by \citet{ref:ParTug04} is shown with 
sky blue lines. The Figure \ref{fig:1} is a Mollweide projection of a part of 
the celestial sphere. 

It can be clearly seen that the apex travels in the same direction with the 
improvement of the sample. As a result, the apex deviated from the results of 
most authors (red lowercase letters in Figure \ref{fig:1}) and became closer to 
the results of \citet{ref:LP}. However, the obtained value of the bulk flow 
velocity $278\,\textrm{km\,s}^{-1}$ is much smaller then that given by 
\citet{ref:LP} and is consistent with $\Lambda$CDM cosmology. Naturally, we 
analyzed the reasons behind this phenomenon. This change of the apex is mostly 
due to the improved measurements of the \mbox{H\,{\sc i}} line widths of the 
galaxies, which were already present in the sample. All such re-measurements 
were double-checked by us and appeared to better match the regression relation. 
It is important to note that the changes in the \mbox{H\,{\sc i}} line width 
have approximately equal probabilities to be positive or negative, thus 
imposing no systematic shift. 

Now let us discuss the sample compilation. This step is very important, because 
it is one of the few manual operations in the routine. Errors at this step can 
lead to underestimation of the quadrupole and higher multipoles due to 
selection.

The new data on radial velocities $V$ and \mbox{H\,{\sc i}} line widths $W$ were 
compared to the data from the previous version of the sample described by 
\citet{ref:APSS10} and published in \citep{ref:arxiv09}. The data about newly 
added galaxies were included automatically. The data, whose difference from the 
old ones either in $V$ or in $W$ was less than a few percent, were considered 
to be the same as before and were not changed. The data, whose difference from 
the old ones exceeded few percent, were considered as alternatives to the old 
data. Thus, we have a problem: if we have several alternative sets of data for 
a galaxy, which one we should choose?

\begin{figure}[tb]
\includegraphics[width=\columnwidth]{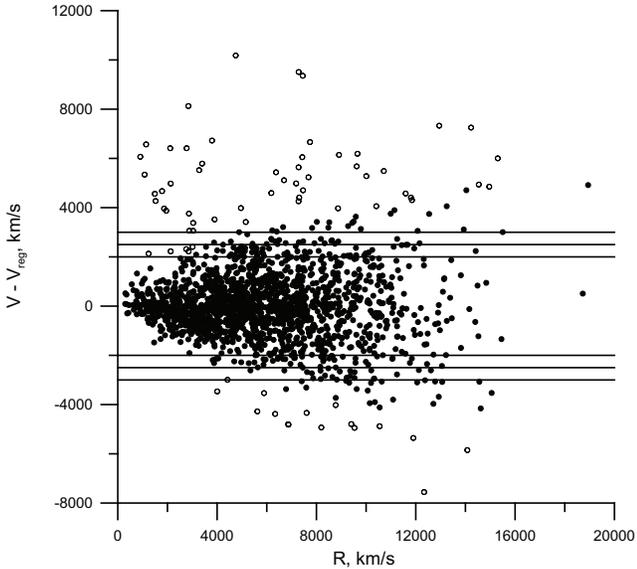}
\caption{The deviation from the D-model vs. the radial distance for all 
galaxies. Horizontal lines show the $2000$, $2500$ and 
$3000\,\textrm{km\,s}^{-1}$ thresholds
}
\label{fig:2}
\end{figure}

We addressed this problem in the following way. In Figure \ref{fig:2} we 
plotted all the galaxies with known \mbox{H\,{\sc i}} line widths. On the horisontal axis we 
plotted the distance according to D-model. On the vertical axis we plotted the
difference between the redshift velocity and the value of the velocity 
following from the D-model at $R_{max}=100h^{-1}\,\textrm{Mpc}$. The galaxies 
which enter the sample are marked with dots, and the rejected galaxies -- with 
circles. Among all possible alternatives we chose those with smallest 
deviations. However, this cannot be done in a straightforward way because the 
value of the deviation depends also on the regression parameters, which are 
sample-specific.

For this reason we calculated the deviations using 3 different regression 
relations. The first one is based on the old sample (it's parameters are given 
in \citep{ref:APSS10}). The second one is based only on the galaxies, which 
enter the regression and have only one set of data. The third one initially 
includes all the candidate data, i.e. all the different sets of data for each 
galaxy were included as separate entries. Then we start rejecting the galaxies 
from the most obvious outliers. At each stage the regression relation is 
recalculated. However, we must stop at some point. Note that we included in the 
sample some galaxies with the deviations up to $3.5\sigma$. Could we reject 
more galaxies and reduce the standard deviation of our sample? Let us 
demonstrate that this will lead to selection.

There are two effects which cause the deviation from the D-model. The first 
one is random scatter, which we will roughly estimate with a normal 
distribution with zero mean and $\sigma^2$ variance. The second one is the 
impact of the quadrupole and higher multipoles, which we will denote as 
$\Delta$. Note that $\Delta$ greatly varies across the sample, taking the 
largest values at the edges of sample and in the vicinity of attractors.

Which galaxies will enter the sample and which will be rejected? If 
we assume the deviation from the D-model to be purely random, i.e. $\Delta = 
0$, we will reject the galaxies whose deviations exceed some threshold. For the 
$2\sigma$ threshold the percentage of rejected galaxies is $2.3\%$ for positive 
deviations and the same for negative ones. For the $2.5\sigma$ threshold these 
percentages are $0.6\%$ in both ways, and for the $3\sigma$ threshold they are 
both less than $0.1\%$. Rejection of these galaxies yields no systematics.

Now let us consider a region of space where the higher multipoles account for a
constant impact of $|\Delta| = \sigma$. The percentage of galaxies in such 
areas can be roughly estimated as $10\%$ of the sample. A non-zero value of 
$\Delta$ will lead to an effective shift of the thresholds by $\Delta$, 
positive for the deviations with the same sign as $\Delta$ and negative in the 
opposite case. This effect causes a selection which leads to an underestimation 
of higher multipoles. In this case the percentages of the rejected galaxies 
will become asymmetric. For the $2\sigma$ threshold the percentages of rejected 
galaxies are $15.8\%$ in the direction of $\Delta$ and less than $0.1\%$ in the 
opposite one. For the $2.5\sigma$ threshold the percentage ion the direction of 
$\Delta$ is $6.5\%$, and for the $3\sigma$ threshold it becomes $2.3\%$. The 
percentages of rejects in the opposite direction for $2.5\sigma$ and $3\sigma$ 
thresholds are negligible. Thus, the introduced selection for this particular 
area is $16\%$ for the $2\sigma$ threshold, $6.5\%$ for the $2.5\sigma$ 
threshold, and $2.3\%$ for the $3\sigma$ threshold. Multiplying this by $10\%$ 
of the sample volume (about 170 galaxies), we get the expected numbers of 
erroneously rejected galaxies: 27 galaxies for the $2\sigma$ threshold, 11 
galaxies for the $2.5\sigma$ threshold and only 4 galaxies for the $3\sigma$ 
threshold.

To verify these simple estimations we considered the subsamples with different 
deviation thresholds. They are shown in Figure \ref{fig:2} with horizontal 
lines corresponding to $2000$, $2500$ and $3000\,\textrm{km\,s}^{-1}$ 
thresholds. Combining these thresholds with DQ- and DQO-models and 4 different 
limits on distance, namely $80 h^{-1}$, $100 h^{-1}$, $140 h^{-1}$ and $170 
h^{-1}\,\textrm{Mpc}$, we get 32 different combinations. For each of them we 
calculated the velocity of the dipolar component as well as the maximal 
$\lambda_1$ and the minimal $\lambda_3$ eigenvalues of the shear tensor 
$\tens{Q}$ (the intermediate eigenvalue is not independent and is equal to 
$\lambda_2=-\lambda_1-\lambda_3$). The maximal dipolar velocity $D$ among all 
possible combinations was equal to $304\,\textrm{km\,s}^{-1}$, which is 
consistent with the $\Lambda$CDM cosmology. The values of the dipolar velocity
and the eigenvalues are given in Table \ref{tbl:1}. One can see that the 
quadrupole significantly drops with the $2500\,\textrm{km\,s}^{-1}$ threshold 
for all subsamples. For the DQ-model with $R_{max} = 80 h^{-1}\,\textrm{Mpc}$ 
subsample it drops already at the $3000\,\textrm{km\,s}^{-1}$ threshold. The 
selection criterion we used to construct the sample appeared to be free from 
such problems.

\begin{figure}[tb]
\includegraphics[width=\columnwidth]{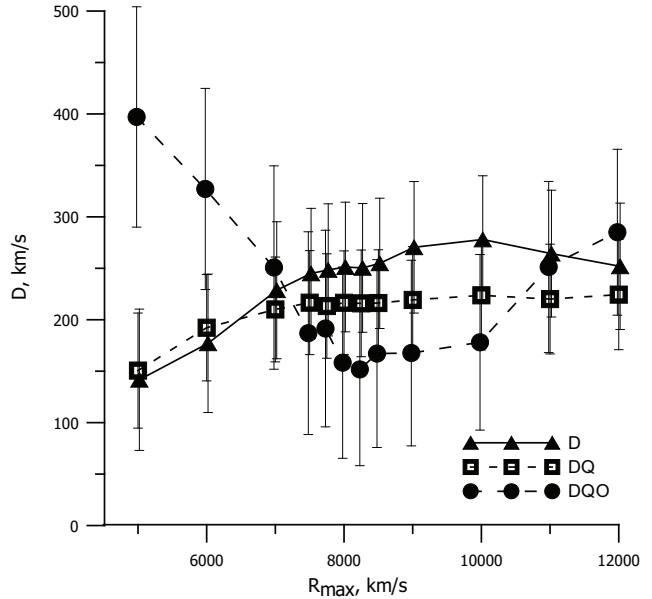}
\caption{The dependence of the norm of the dipolar velocity component on the 
sample depth $R_{max}$ in the framework of D, DQ and DQO models}
\label{fig:3}
\end{figure}

We obtained the norm of the dipolar velocity component for different models and 
different subsamples. In Figure \ref{fig:3} we plotted its dependence on the 
sample depth $R_{max}$ in the framework of D, DQ and DQO models. In the 
simplest D-model the norm grows up to the distance $100h^{-1}\,\textrm{Mpc}$, 
in the DQ-model it is almost constant beyond $70h^{-1}\,\textrm{Mpc}$, and in 
the DQO-model it has a minimum at about $82.5h^{-1}\,\textrm{Mpc}$ (the 
minimal norm is non-zero at $1.5\sigma$ confidence level). For 
$R_{max}>120h^{-1}\,\textrm{Mpc}$ these values are almost constant due to 
the small number of more distant galaxies in our sample (47 out of 1661).

These results are obtained for unit weights of all data points. Let us see how 
these results will be affected by weighting. Let us choose the weights, which 
decrease with distance inverse proportional to $\Delta^2(R)$, where $\Delta(R)$ 
is a maximum deviation of the velocity from the D-model at the distance R. 
After calculating the norm of the dipolar velocity component for the subsamples 
limited at 80, 100 and $120h^{-1}\,\textrm{Mpc}$ we obtain for the 
D-model 243, 251, and $243\,\textrm{km\,s}^{-1}$ respectively, for the DQ-model 
245, 248, and $244\,\textrm{km\,s}^{-1}$, and for the DQO-model 250, 262, and 
$280\,\textrm{km\,s}^{-1}$. All these values do not differ essentially from 
those with unit weighting and still are consistent with $\Lambda$CDM cosmology.

\begin{table*}
\caption{The impact of the deviation threshold on the dipolar velocity and the maximal and minimal eigenvalues of the quadrupole}
\begin{tabular}{|c|cccc|cccc|cccc|}
\hline
\multirow{2}{*}{Max deviation, $\textrm{km\,s}^{-1}$}&\multicolumn{4}{c|}{$D,\,\textrm{km\,s}^{-1}$}&\multicolumn{4}{c|}{$\lambda_1,\,\%$}&\multicolumn{4}{c|}{$\lambda_3,\,\%$}\\
\cline{2-13}
&Full&3000&2500&2000&Full&3000&2500&2000&Full&3000&2500&2000\\
\hline
$R_{max},\,h^{-1}\,\textrm{Mpc}$&\multicolumn{12}{c|}{DQ-model}\\
\hline
$ 80$&$217$&$197$&$186$&$195$&$7.4$&$3.1$&$2.5$&$3.5$&$-5.8$&$-3.7$&$-4.0$&$-3.3$\\
$100$&$224$&$213$&$195$&$227$&$6.2$&$6.2$&$3.5$&$2.5$&$-4.2$&$-4.0$&$-3.3$&$-3.1$\\
$140$&$233$&$222$&$196$&$225$&$3.7$&$3.8$&$2.4$&$2.8$&$-3.9$&$-4.5$&$-3.4$&$-4.0$\\
$170$&$228$&$225$&$196$&$227$&$4.0$&$3.9$&$2.4$&$2.9$&$-4.4$&$-4.5$&$-3.4$&$-4.2$\\
\hline
$R_{max},\,h^{-1}\,\textrm{Mpc}$&\multicolumn{12}{c|}{DQO-model}\\
\hline
$ 80$&$158$&$304$&$269$&$258$&$7.1$&$7.0$&$5.6$&$3.5$&$-5.6$&$-6.5$&$-6.3$&$-4.0$\\
$100$&$178$&$212$&$283$&$275$&$7.3$&$7.1$&$4.7$&$2.9$&$-4.8$&$-4.8$&$-4.4$&$-4.0$\\
$140$&$292$&$261$&$269$&$236$&$7.0$&$7.3$&$5.2$&$4.3$&$-6.5$&$-6.5$&$-5.5$&$-6.0$\\
$170$&$304$&$260$&$269$&$237$&$7.1$&$7.3$&$5.2$&$4.3$&$-6.7$&$-6.4$&$-5.5$&$-6.2$\\
\hline
\end{tabular}
\label{tbl:1}
\end{table*}

Additionally, we analyzed the impact of the 16-pole and 32-pole on the velocity 
field. The goal of this analysis was to analyze how the inclusion of higher 
multipoles affects the dipole, the quadrupole and the octupole. The statistical 
significance of both multipoles exceeds $99.9\%$ according to F-test and the 
inclusion of the 16-pole reduces the RMS error by $4.5\%$. The 
semi-relativistic DQOX-model gives a much better apex as compared to the 
semi-relativistic DQO-model and the value of the dipolar velocity 
$236\,\textrm{km\,s}^{-1}$ is also reasonable. Thus, the semi-relativistic 
DQOX-model can be used for deep samples. The DQOXT-model contains too much 
parameters and it is too early to be considered unless the sample volume and 
accuracy is essentially improved. The maximal dipolar velocity in these models 
appeared to be equal to $342\,\textrm{km\,s}^{-1}$ in the DQOXT-model with 
$R_{max} = 80 h^{-1}\,\textrm{Mpc}$, which is marginally consistent with 
$\Lambda$CDM cosmology. 

In Figure \ref{fig:1} we plotted the apices of the dipolar velocity in 
different models of motion. The letter D denotes the D-model, Q -- the 
DQ-model, O -- the DQO-model, X -- the DQOX model and T -- the DQOXT-model. For 
DQO-, DQOX- and DQOXT-models both the non-relativistic and the 
semi-relativistic versions are shown with black and blue colour respectively. 
One can see that the non-relativistic DQO-model gives the apex closest to the 
results of most authors. This is consistent with our earlier results 
\citep{ref:ParPar08,ref:APSS11} of Monte Carlo simulations, which demonstrate 
that the DQO-model gives the best fit to the data. However, the apex in the 
semi-relativistic DQO-model is way beyond all reasonable confidence boundaries. 
This is due to the low value of the dipolar velocity in this model, which 
yields large errors in the determination of the apex. 

For reference we give the parameters of the non-relativistic DQO-model at 
$R_{max}=100h^{-1}\,\textrm{Mpc}$. The residual mean square error is $\sigma 
= 1112\,\textrm{km\,s}^{-1}$.

The coefficients of the generalised Tully-Fisher relation are
\begin{equation}\label{eqn:ctf}
\begin{array}{l}
C_1 = 18.0\pm 1.3, C_2 = 1.59\pm 1.8,\\
C_3 = -0.27\pm 0.11, C_4 = 6.5\pm 1.1,\\
C_5 = (-7.2\pm 1.1)\cdot 10^{-3}, C_6 = -919\pm 82.
\end{array}
\end{equation}

The dipolar velocity is equal to $D = 178\,\textrm{km\,s}^{-1}$ and points to 
the apex with the galactic coordinates $l = 303\degr$, $b = +11\degr$. The 
components of the dipolar velocity are equal to $D_z = 35\pm 
74\,\textrm{km\,s}^{-1}$, $D_x = 94\pm 95\,\textrm{km\,s}^{-1}$, $D_y = -147\pm 
94\,\textrm{km\,s}^{-1}$. \citet{ref:FWH10} using a model similar to DQO on the 
sample with the same depth $100 h^{-1}\,\textrm{Mpc}$ obtained the dipolar 
velocity $D = 416\pm 78\,\textrm{km\,s}^{-1}$ pointed towards $l = 282\pm 
11\degr$, $b = +6\pm 6\degr$. 

The quadrupolar component is described by its 5 irreducible components:
\begin{equation}\label{eqn:ocomp}
\begin{array}{l}
q_1 = (7.2\pm 1.5)\%,
q_2 = (-2.7\pm 1.6)\%,\\
q_3 = (-0.8\pm 2.0)\%,
q_4 = (1.9\pm 2.4)\%,\\
q_5 = (-1.6\pm 2.6)\%.
\end{array}
\end{equation}
It can be also represented by its eigenvalues $\lambda_1 = (7.27\pm 1.54)\%$, 
$\lambda_2 = (-2.43\pm 1.46)\%$, $\lambda_3 = (-4.84\pm 1.47)\%$ and 
eigenvectors pointing respectively to $l = 118\degr$, $b = +85\degr$ (Canes 
Venatici); $l = 341\degr$, $b = +4\degr$ (Scorpius/Norma); and $l = 71\degr$, 
$b = -4\degr$ (Cygnus/Volans) and in the opposite directions: Sculptor, Auriga 
and Centaurus/Vela/Carina. The direction of the eigenvector corresponding to 
the maximum eigenvalue is close to the Supergalactic plane and the direction 
corresponding to the intermediate eigenvalue is not far from the direction 
towards the Great Attractor. These directions are marked in Figure \ref{fig:1} 
with numbers 1 (maximum), 2 (intermediate) and 3 (minimum). It is interesting 
to compare these values with those obtained by \citet{ref:CHTG}, who considered 
a sample of 1797 galaxies within $30 h^{-1}\,\textrm{Mpc}$. Their directions of 
the eigenvectors demonstrated good agreement with our results for $100 
h^{-1}\,\textrm{Mpc}$. However, our results for $30 h^{-1}\,\textrm{Mpc}$ were 
in not so good agreement due to a small number of galaxies in this subsample. 

The octupolar component is described by 10 irreducible components:
\begin{equation}\label{eqn:ocomp}
\begin{array}{l}
P_1 = (2.2\pm 2.1)\cdot 10^{-6}\,\textrm{s\,km}^{-1},\\
P_2 = (1.0\pm 2.5)\cdot 10^{-6}\,\textrm{s\,km}^{-1},\\
P_3 = (0.4\pm 3.1)\cdot 10^{-6}\,\textrm{s\,km}^{-1},\\
o_1 = (3.0\pm 1.5)\cdot 10^{-6}\,\textrm{s\,km}^{-1},\\
o_2 = (1.0\pm 1.8)\cdot 10^{-6}\,\textrm{s\,km}^{-1},\\
o_3 = (6.7\pm 1.8)\cdot 10^{-6}\,\textrm{s\,km}^{-1},\\
o_4 = (-2.9\pm 2.2)\cdot 10^{-6}\,\textrm{s\,km}^{-1},\\
o_5 = (5.5\pm 2.0)\cdot 10^{-6}\,\textrm{s\,km}^{-1},\\
o_6 = (-2.0\pm 2.1)\cdot 10^{-6}\,\textrm{s\,km}^{-1},\\
o_7 = (22.0\pm 7.3)\cdot 10^{-6}\,\textrm{s\,km}^{-1}.
\end{array}
\end{equation}
The trace vector $\vec{P}$, which is a part of the reduced octupole tensor and shows 
how the dipolar component changes with distance \citep{ref:Par01}, points 
towards the direction $l = 23\degr$, $b = +64\degr$ (Bo\"otes).

\begin{figure}[tb]
\includegraphics[width=\columnwidth]{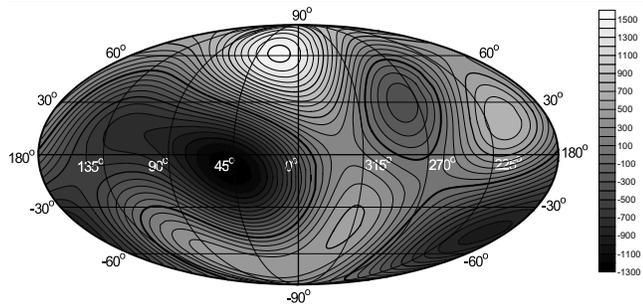}
\caption{The radial component of the velocity field calculated in the 
non-relativistic DQO-model at $R = 80h^{-1}\,\textrm{Mpc}$}
\label{fig:4}
\end{figure}

In Figure \ref{fig:4} we plotted the radial component of the velocity field 
calculated in the non-relativistic DQO-model with above parameters at $R = 80 
h^{-1}\,\textrm{Mpc}$. The global minimum lies in the direction opposite to the 
apex and the global maximum matches the direction of the maximum eigenvector of 
the quadrupole component. This can explain the shift of the apex in the D-model 
towards the latter. 

Note that each multipole of order $n\ge 2$, i.e. quadrupole and higher, can be 
represented in a reduced form, in which the multipoles of orders $n-2$, $n-4$ 
and so on are separated. However, it is important to realize that these lower 
order multipoles have the same dependence on the radial distance as the initial 
one. We use the reduced form for the quadrupole, separating the Hubble constant 
$H$, and for the octupole, separating the trace vector $\vec{P}$, which 
describes the change of the bulk flow velocity with distance (see 
\citep{ref:Par01} for details). We do not use the reduced form for the 16-pole 
and the 32-pole, because their separated lower-order parts do not have obvious 
physical meaning. In addition, since these lower-order parts have a different 
dependence on the distance than the quadrupole and the octupole, there is no 
interference between them and both their reduced and full forms can be used 
interchangeably.

\section{Conclusion}
We described a new sample of RFGC galaxies intended for the study of 
large-scale cosmic flows. A preliminary verification with the methods described 
in our previous articles has shown that this new sample can be used as a basis 
for the study of collective motions of galaxies. We plan to modify our routine 
to reconstruct the distribution of matter density simultaneously with the 
parameters of the model of collective velocity field. In this article we 
applied the old routine to the new sample to verify its quality. The obtained 
results show principal agreement with the previous ones. It is essential that 
in all considered cases we obtained the values of the dipolar velocity, which 
are consistent with the $\Lambda$CDM cosmology. Once again, it was demonstrated 
that the samples with significant depth can not be considered in the framework 
of a simple D-model featuring only Hubble expansion and the bulk flow. For our 
sample with the depth $80-100 h^{-1}\,\textrm{Mpc}$ the optimal choice is the 
DQO-model, which also takes into account the cosmic shear and the octupolar 
components. For the first time, higher-order multipoles were considered and 
appeared to be statistically significant. The inclusion of the 16-pole was 
beneficial for the semi-relativistic model and improved the obtained results.

\acknowledgments

This research has made use of the NASA/IPAC Extragalactic Database (NED) which
is operated by the Jet Propulsion Laboratory, California Institute of
Technology, under contract with the National Aeronautics and Space
Administration.

\end{document}